\documentclass[%
 reprint,
superscriptaddress,
 amsmath,amssymb,
 aps,
prb,
citeautoscript,
]{revtex4-2}

\usepackage{graphicx}
\usepackage{dcolumn}
\usepackage{bm}
\usepackage{siunitx}
\usepackage{braket}
\usepackage{tikz}
\usetikzlibrary{quantikz2}
\usepackage{floatrow} 
\usepackage{multirow} 
\usepackage{csquotes}
\usepackage{booktabs}
\usepackage{algorithm}
\usepackage{algorithmic}
\usepackage{orcidlink}  
\usepackage[caption=false]{subfig}   
\usepackage{mleftright} 

\newcommand{\figref}[1]{\mbox{Fig.~\ref{#1}}}

\newcommand{\secref}[1]{\mbox{Sec.~\ref{#1}}}

\newcommand{\appref}[1]{\mbox{Appendix~\ref{#1}}}
\renewcommand{\eqref}[1]{\mbox{Eq.~(\ref{#1})}}
\newcommand{\figpanel}[2]{Fig.~\hyperref[#1]{\ref*{#1}(#2)}} 

\newcommand{\figpanels}[3]{Fig.~\hyperref[#1]{\ref*{#1}(#2)--(#3)}} 
\newcommand{\figpanelNoPrefix}[2]{\hyperref[#1]{\ref*{#1}(#2)}} 
\newcommand{\figpanelsNoPrefix}[3]{\hyperref[#1]{\ref*{#1}(#2)--(#3)}} 

\begin{document}

\preprint{APS/123-QED}

\title{Generative flow-based warm start of the variational quantum eigensolver}  

\author{Hang Zou}
\affiliation{Department of Computer Science and Engineering, Chalmers University of Technology and University of Gothenburg, 41296 Gothenburg, Sweden} 

\author{Martin Rahm}
\affiliation{Department of Chemistry and Chemical Engineering, Chalmers University of Technology, 41296 Gothenburg, Sweden} 

\author{Anton Frisk Kockum}
\affiliation{Department of Microtechnology and Nanoscience, Chalmers University of Technology, 41296 Gothenburg, Sweden} 

\author{Simon Olsson}
\email{simonols@chalmers.se}
\affiliation{Department of Computer Science and Engineering, Chalmers University of Technology and University of Gothenburg, 41296 Gothenburg, Sweden} 

\date{\today}


\begin{abstract}

Hybrid quantum-classical algorithms like the variational quantum eigensolver (VQE) show promise for quantum simulations on near-term quantum devices, but are often limited by complex objective functions and expensive optimization procedures. Here, we propose Flow-VQE, a generative framework leveraging conditional normalizing flows with parameterized quantum circuits to efficiently generate high-quality variational parameters. By embedding a generative model into the VQE optimization loop through preference-based training, Flow-VQE enables quantum gradient-free optimization and offers a systematic approach for parameter transfer, accelerating convergence across related problems through warm-started optimization. We compare Flow-VQE to a number of standard benchmarks through numerical simulations on molecular systems, including hydrogen chains, water, ammonia, and benzene. We find that Flow-VQE outperforms baseline optimization algorithms, achieving computational accuracy with fewer circuit evaluations (improvements range from modest to more than two orders of magnitude) and, when used to warm-start the optimization of new systems, accelerates subsequent fine-tuning by up to 50-fold compared with Hartree--Fock initialization. Therefore, we believe Flow-VQE can become a pragmatic and versatile paradigm for leveraging generative modeling to reduce the costs of variational quantum algorithms.
 
\end{abstract}

\maketitle



\section{Introduction}

The advent of noisy intermediate-scale quantum (NISQ)~\cite{preskill2018quantum} computers~\cite{Kim2023, Bluvstein2024, Acharya2025} has spurred significant interest in hybrid quantum-classical algorithms~\cite{Montanaro2016, cerezo2021variational, Cerezo2022, Dalzell2023} that can leverage current quantum hardware capabilities while mitigating their limitations using conventional computation. Among them, the variational quantum eigensolver (VQE)~\cite{peruzzo2014variational, cerezo2021variational} is a widely adopted approach that uses parameterized quantum circuits, together with classical optimization of the parameters, to approximate the ground state of a many-body Hamiltonian. Although VQE does not overcome the worst-case computational complexity of quantum simulation --- unstructured variational optimization is NP-hard~\cite{bittel2021training} --- it remains a promising heuristic framework, combining physically motivated ansätze, initialization schemes, and optimization strategies, whose practical value warrants continued investigation~\cite{cerezo2021variational, zimborás2025myths}. Here, we propose a machine learning approach for learning the distribution of good variational parameters and producing promising initial parameter guesses, taking into account physical information about the system.

A major challenge of variational quantum algorithms is that the optimization landscape is typically highly nonconvex and noisy. Phenomena such as barren plateaus~\cite{mcclean2018barren, larocca2024review} and the proliferation of local minima~\cite{anschuetz2022quantum} substantially hinder convergence, often necessitating a large number of quantum circuit evaluations to achieve acceptable accuracy. Gradient-based optimization methods incur considerable overhead from quantum gradient estimation~\cite{sweke2020stochastic, stokes2020quantum, fitzek2024optimizing}, whereas gradient-free methods typically require even more queries and scale poorly with increasing parameter dimensionality~\cite{gacon2021simultaneous, wanner2025variational}. Furthermore, conventional optimization procedures tend to focus on isolated optimal parameter configurations, and thus may not fully exploit the underlying structure of the parameter landscape, which can limit their capacity for transferring knowledge between related tasks.

One promising strategy to address these challenges is the use of warm-start techniques, which aim to initialize optimization with informed guesses positioned within advantageous regions of the optimization landscape, rather than random configurations. One class of methods focuses on encoding prior knowledge into initial quantum states, for example, by leveraging chemistry-inspired solutions~\cite{tubman2018postponing,fomichev2024initial} or biased state preparation informed by classical algorithms~\cite{zhang2021mutual, truger2024warm, dborin2022matrix, rudolph2023synergistic}, thereby steering the search toward physically meaningful regions of the Hilbert space.
Complementary strategies aim to initialize variational parameters through parameter transfer, reusing parameters obtained from previously solved or structurally related problems to increase the likelihood that new instances are initialized within a favorable training region~\cite{skogh2023accelerating, liu2023mitigating, mele2022avoiding, rohe2025accelerated, lyngfelt2025symmetry, puig2025variational}. 

Building on the success of heuristic warm-start methods, recent efforts have increasingly explored machine learning-based approaches that aim to enhance flexibility and applicability across diverse problem settings. Models trained on precomputed quantum data have been employed to produce effective initial parameters through supervised learning~\cite{tao2022exploring, yao2024machine} and generative modeling~\cite{ceroni2022generating, zhang2025diffusion}, enabling generalization to unseen problem instances while supporting extremely fast sampling during inference.
In parallel, emerging data-free paradigms integrate machine learning models directly into the quantum optimization loop to provide dynamic guidance and modify the cost landscape~\cite{miao2024neural, mesman2024nn, rivera2021avoiding, zhang2022variational}.
Extending this idea, meta-learning frameworks train across a distribution of tasks to acquire generalizable optimization behaviors. Rather than fitting solutions, meta-learners directly interact with optimization trajectories during training, progressively distilling shared initialization strategies from multiple tasks to enable rapid adaptation to new structures~\cite{verdon2019learning, cervera2021meta, sauvage2021flip, kamata2025molecular, chang2025accelerating}. 

In this article, we introduce Flow-VQE, a probabilistic framework that learns the distribution of ``high-quality'' variational parameters  that yield low-energy quantum states. By modeling this distribution, Flow-VQE equips VQE algorithms with an adaptable, learnable prior, which allows for the one-shot generation of effective initial solutions and significantly diminishes the need for costly iterative optimization from scratch.  Flow-VQE leverages flow-based generative models~\cite{tabak2010density,rezende2015variational, kobyzev2020normalizing, papamakarios2021normalizing} to explicitly model conditional probability distributions over variational parameters, conditioned on relevant contextual information of quantum systems. We develop a preference-based optimization approach for Flow-VQE, enabling efficient training while circumventing quantum gradient calculations. In doing so, the generative model replaces the classical optimizer, turning direct gradient-query interactions into a sampling-based dialogue with the quantum circuit. Additionally, training the model on a suite of multi-objective tasks enables it to acquire meta-initialized heuristics for new systems.
 
We empirically validate Flow-VQE through state-vector simulation experiments on various quantum chemical systems. In single-molecule tasks, Flow-VQE reaches computational accuracy with up to two orders of magnitude fewer circuit evaluations than gradient-descent algorithms. When used as a warm start for subsequent fine-tuning, it delivers as much as 50-fold acceleration, while the total training overhead remains no greater than that required to optimize five molecules by conventional methods. Our results comprehensively demonstrate that Flow-VQE significantly reduces quantum circuit evaluations and reliably generates high-quality variational parameters for initialization, paving the way for its establishment as a foundational technique for future warm-starts in variational quantum algorithms.

The remainder of this article is organized as follows, to provide the conceptual and empirical ground for Flow-VQE. In \secref{sec:background}, we review the essential ingredients of the VQE and flow-based generative models. In \secref{sec:methodology}, we detail the Flow‑VQE motivation, workflow, and optimization methods that enable training without quantum gradients. In \secref{sec:numerical}, we describe simulation experiments and molecular benchmarks used to quantify performance. In \secref{sec:results}, we analyze the resulting cost savings in both single‑task optimization and generative warm starts. We discuss practical limitations and avenues for future improvement in \secref{sec:future} and conclude in \secref{sec:conclusion}.


\section{Background}
\label{sec:background}
 
In this section, we first provide a concise recap of the VQE framework. We then review normalizing flows, showing how to use them for data-distribution learning and sample generation. Finally, we introduce a benchmark parameter-transfer technique that leverages geometric proximity to accelerate VQE convergence.


\subsection{Variational quantum eigensolver}

The VQE algorithm is a hybrid quantum-classical optimization framework designed to approximate the ground-state energies of complex quantum systems~\cite{peruzzo2014variational, mcclean2016theory, cerezo2021variational}. Given a Hamiltonian operator $\hat H$ defined on an $n$-qudit ($D$-level quantum system) Hilbert space $\mathcal{H}\in\mathbb{C}^{D^n\times D^n}$, the VQE seeks to determine its minimal eigenvalue $E_0$, defined as:
\begin{equation}
E_0 = \min_{\psi \in \mathcal{H}, |\psi|=1} \langle\psi|\hat{H}|\psi\rangle.
\end{equation}  
Although the VQE can in principle be applied to arbitrary multilevel quantum systems, in this work we focus on the case $D=2$, i.e., standard qubit circuits.
To do so, the VQE involves preparing a trial quantum state using a variational ansatz $U(\boldsymbol{\theta})$: 
\begin{equation}
    \psi(\boldsymbol{\theta}) = U(\boldsymbol{\theta}) \ket 0 ^{\otimes n},
\end{equation}
where $\boldsymbol{\theta}=(\theta_1,\theta_2,\ldots, \theta_d) \in \mathbb{R}^d$ is a vector of trainable parameters, and  $U(\boldsymbol{\theta}): \mathbb{R}^d \rightarrow \mathcal{U}(2^n)$ is a smooth mapping from the parameter space to the unitary group acting on $\mathcal{H}$. 
This formulation rephrases the eigenvalue problem as a variational optimization problem in $\boldsymbol{\theta}$:
\begin{equation}
\boldsymbol{\theta}^* = \underset{\boldsymbol{\theta} \in \mathbb{R}^d}{\arg\min}~\mathcal{L}(\boldsymbol{\theta}) ,
\end{equation}
where $\mathcal{L}(\boldsymbol{\theta}):= \langle\psi(\boldsymbol{\theta})|\hat{H}|\psi(\boldsymbol{\theta})\rangle$ is the variational energy functional (expected energy of the trial state).

The optimization proceeds as a quantum-classical loop: a classical optimizer suggests parameter updates, the quantum device evaluates the objective function $\mathcal{L}(\boldsymbol{\theta})$ and its gradients with respect to these parameters, and then the results are iteratively fed back until convergence. Among various optimization strategies, gradient-based methods such as gradient descent are commonly employed. These methods require quantum gradient information, e.g., using the parameter-shift rule~\cite{mitarai2018quantum}. This gradient estimation has a computational complexity of $\mathcal{O}(d)$ per iteration, where $d$ is the number of parameters in the ansatz, making it substantially more expensive than classical backpropagation, which, in contrast, computes all parameter derivatives simultaneously in a single pass. Alternatively, gradient-free methods can be developed under different design principles and have received significant attention~\cite{lavrijsen2020classical, chivilikhin2020mog, jager2024fast, gacon2021simultaneous, wanner2025variational}. These approaches avoid explicit gradient computation and can potentially be more practical in scenarios where evaluating quantum gradients is costly or infeasible, such as under hardware noise or in the presence of non-differentiable objectives.


\subsection{Normalizing flows}

Normalizing flows (NFs)~\cite{tabak2010density, rezende2015variational, papamakarios2021normalizing} are a class of generative models predicated on the principle of invertible maps between probability densities. To see how NFs work, consider a latent vector $\mathbf{z} \in \mathbb{R}^d$ with an associated simple base distribution $p_Z(\mathbf{z})$, typically a multivariate Gaussian distribution $\mathcal{N}(\boldsymbol{\mu}, \boldsymbol{\Sigma})$. The fundamental objective is to construct a bijective and differentiable ({\it diffeomorphic}) map $f_{\boldsymbol{{\boldsymbol{\tau}}}}: \mathbb{R}^d \rightarrow \mathbb{R}^d$ with model parameters $\boldsymbol{{\boldsymbol{\tau}}} \in \mathbb{R}^p$, which transforms $p_Z(\mathbf{z})$ into a modeling distribution $p_X(\mathbf{x})$ that approximates a complex, unknown data distribution.  

The transformation is governed by the change-of-variables formula:
\begin{equation}
p_X(\mathbf{x})=p_Z\mleft(\mathbf{z}\mright)\mleft|\operatorname{det}\frac{\partial \mathbf{z}}{\partial \mathbf{x}}\mright|=p_Z\mleft(f_{\boldsymbol{\tau}}^{-1}[\mathbf{x}]\mright)\mleft|\operatorname{det}\mathbf{J}\mleft({f_{\boldsymbol{\tau}}^{-1}}[\mathbf{x}]\mright)\mright|,
\end{equation}  
where $\mathbf{J}\mleft({f_{\boldsymbol{\tau}}^{-1}}[\mathbf{x}]\mright)$ denotes the Jacobian matrix of the inverse mapping $f_{\boldsymbol{\tau}}^{-1}$ evaluated at point $\mathbf{x}$.
To model complex distributions tractably, NFs employ compositional transformations:  $f_{\boldsymbol{\tau}}=f_{{\boldsymbol{\tau}}_K} \circ f_{{\boldsymbol{\tau}}_{K-1}} \circ \cdots \circ f_{{\boldsymbol{\tau}}_1}$, where $K$ is the number of transformations.

This compositional design allows the overall model to be highly expressive, while keeping each individual Jacobian manageable for likelihood evaluation. However, computing the log-likelihood remains computationally intensive due to the need to evaluate Jacobian determinants at each layer, resulting in a time complexity of $\mathcal{O}(Kd^3)$ for general diffeomorphic maps. To alleviate this cost, modern flow architectures are often constructed to have sparse Jacobians, where algorithms with more favorable time complexity for computing the Jacobian determinant are possible. Some examples include affine coupling flows~\cite{dinh2016density,kingma2018glow}, autoregressive flows~\cite{kingma2016improved,papamakarios2017masked}, and spline-based flows~\cite{muller2019neural,durkan2019cubic}. In contrast, the generation process exhibits remarkable efficiency, requiring only $\mathcal{O}(Kd)$ computational complexity, as it involves mere function evaluations without determinant calculations, enabling rapid sample synthesis.

 
\subsubsection{Forward training process} 

In the training phase, NFs apply the inverse transformation to map each data point from the empirical data set $\mathcal{D}=\{\mathbf{x}_i\in \mathbb{R}^d\}_{i=1}^N$ into the latent space via $\mathbf{z}=f_{\boldsymbol{\tau}}^{-1}(\mathbf{x})$. The corresponding log-likelihood can be evaluated by 
\begin{equation}
\label{eq:logp}
\log p_X\mleft(\mathbf{x} ; \boldsymbol{\tau}\mright)=\log p_Z\mleft(\mathbf{z}\mright)+\sum_{j=1}^K\log \mleft|\operatorname{det} \mathbf{J}\mleft(f_{{\boldsymbol{\tau}}_j}^{-1}\mleft[\mathbf{x}\mright]\mright)\mright|.
\end{equation}
The model parameters $\boldsymbol{\tau}$ are trained via maximum likelihood estimation (MLE) over the target data:
\begin{equation}
\label{eq:nll}
\mathcal{L}(\boldsymbol{\tau}; \mathcal{D})=-\mathbb{E}_{\mathbf{x} \sim \mathcal{D}}\mleft[\log p_X(\mathbf{x}; \boldsymbol{\tau})\mright].
\end{equation}


\subsubsection{Backward generation process}

At inference time, we sample $\mathbf{z} \sim p_Z(\mathbf{z})$ from the base distribution and execute the forward transformation $\mathbf{x}=f_{\boldsymbol{\tau}} (\mathbf{z})$. 
For compositional flows, this entails a sequential application of transformations:
\begin{equation}
\mathbf{z} \sim p_Z(\mathbf{z}) \xrightarrow{f_{\boldsymbol{\tau}_1}} \mathbf{h}^{(1)} \xrightarrow{f_{\boldsymbol{\tau}_2}} \mathbf{h}^{(2)} \xrightarrow{~\ldots~}\xrightarrow{ f_{\boldsymbol{\tau}_K}} \mathbf{x},
\end{equation} 
where $\mathbf{h}^{(j)}$ denotes the intermediate output after the $j$-th transformation. 


\subsection{Parameter transfer in the variational quantum eigensolver}

Traditional parameter transfer (PT) accelerates the VQE algorithm by reusing optimized parameters from structurally adjacent problem instances~\cite{skogh2023accelerating,rohe2025accelerated}. Let $\boldsymbol{\theta}_k^\ast$ denote the optimized variational parameters for a VQE task $\mathcal{T}_k$. The PT strategy initializes the subsequent task $\mathcal{T}_{k+1}$ using $\boldsymbol{\theta}_k^\ast$ as a warm start: $\boldsymbol{\theta}_{k+1}^{(0)}:=\boldsymbol{\theta}_k^\ast$. The selection of transferable task pairs typically relies on certain heuristics that aim to preserve underlying similarity between instances. 
One example is using small geometric perturbations~\cite{skogh2023accelerating}, such as those along a potential energy surface (PES), where configurations differ by a small Euclidean distance, e.g., $\mleft\|\boldsymbol R_{k+1}-\boldsymbol R_k\mright\| \sim 0.1~ \mathrm{\AA}$. In the absence of further task-specific guidance, such perturbations are commonly assumed --- with high empirical confidence --- to keep $\boldsymbol{\theta}_k^\ast$  within the attraction basin of $\mathcal{T}_{k+1}$, thereby enabling efficient parameter reuse.


\section{Methodology}
\label{sec:methodology}

We now introduce the Flow-VQE framework as a surrogate model for VQE training, in which conditional normalizing flows can perform one-shot generation of candidate circuit parameters. We then discuss the standard policy-gradient method as a general training paradigm, highlighting their limitations in low-signal, resource-constrained quantum computing application. To overcome these challenges, we introduce a preference-based optimization scheme that enables efficient training of Flow-VQE with few sampling overhead.


\subsection{Overview of Flow-VQE}

\begin{figure*}
    \centering \includegraphics[width=\linewidth]{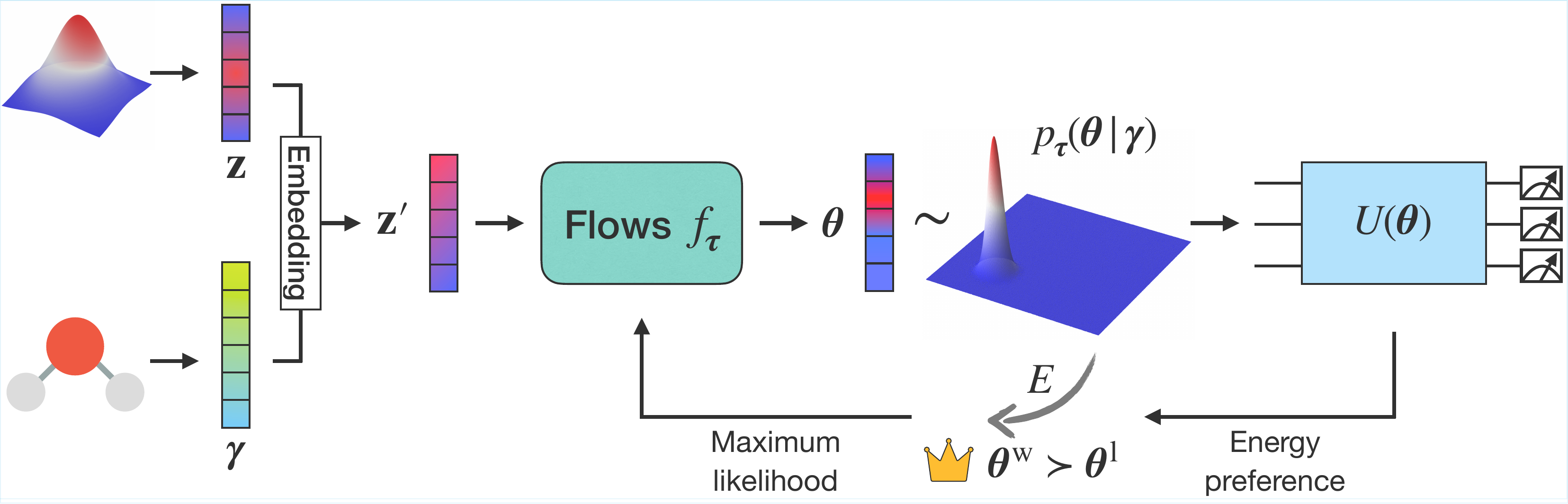} 
    \caption{Schematic overview of the Flow-VQE framework. Latent samples $\mathbf{z}$ conditioned on the molecular context $\boldsymbol{\gamma}$ are transformed into variational parameters $\boldsymbol{\theta}$ via normalizing flows and evaluated by quantum circuits. Preference comparisons identify low-energy winners, which are retained in a buffer and used to update the model through maximum likelihood training.}   
    \label{fig:FIG1}
\end{figure*}

Generative modeling aims to maximize the likelihood of target data, a goal seemingly ill-suited to VQE optimization, since it presupposes access to optimal parameter samples --- resources generally unavailable in this context. Nonetheless, as we will show, this challenge can be overcome by leveraging the flexibility of NFs: they can be trained based on self-sampled data, with the training process guided by external quality evaluations. A compelling precedent for this paradigm is Boltzmann generators~\cite{No2019}, which utilize NFs to learn complex equilibrium distributions by using a physical energy function to guide training in the absence of explicit training data. Inspired by this approach, we treat the VQE energy expectation as an implicit surrogate for distributional fitness, thereby guiding the flow model to concentrate probability density in low-energy regions of parameter space. Here, the term ``implicit'' indicates that the model is trained not through labels or target parameters, but rather through indirect feedback provided by measured energy expectations. The main idea of Flow-VQE is as follows (see \figref{fig:FIG1}):

\begin{enumerate}
    \item[] \textbf{Latent sampling:} Draw samples $\mathbf{z} \sim p(\mathbf{z})$ from a simple, analytically tractable prior $p(\mathbf{z})$.
    \item[] \textbf{Conditional transformation:} Use an invertible neural network $f_{\boldsymbol{\tau}}$ to map $\mathbf{z}$ into variational parameters $\boldsymbol{\theta} = f_{\boldsymbol{\tau}}(\mathbf{z}; \boldsymbol{\gamma})$, conditioned on the problem-specific context $\boldsymbol{\gamma}$, yielding samples $\boldsymbol{\theta} \sim p_{\boldsymbol{\tau}}(\boldsymbol{\theta} | \boldsymbol{\gamma})$. 
    \item[] \textbf{Likelihood evaluation:} Explicitly determine the conditional likelihood $p_{\boldsymbol{\tau}}(\boldsymbol{\theta} | \boldsymbol{\gamma})$ according to \eqref{eq:logp}. 
    \item[] \textbf{Energy measurement:} Run parameterized quantum circuits with $\boldsymbol{\theta}$ to measure the energy expectation value $\langle \psi(\boldsymbol{\theta}) | \hat H(\boldsymbol{\gamma}) | \psi(\boldsymbol{\theta}) \rangle$, providing the implicit training signal for the generative model.
\end{enumerate}

The general objective for Flow-VQE is to minimize the expected energy across problem instances:
\begin{equation}\begin{aligned}
    \boldsymbol{\tau}^* &= \arg \min_{\boldsymbol{\tau}} \mathcal{L}(\boldsymbol \tau) \\
    &=\arg \min_{\boldsymbol{\tau}} \mathbb{E}_{\boldsymbol{\gamma}} \mleft[ \mathbb{E}_{\boldsymbol{\theta} \sim p_{\boldsymbol{\tau}}(\boldsymbol{\theta} | \boldsymbol{\gamma})} \mleft[ \langle \psi(\boldsymbol{\theta}) | \hat H(\boldsymbol{\gamma}) | \psi(\boldsymbol{\theta}) \rangle \mright] \mright].
    \label{eq:8}
\end{aligned} 
\end{equation} 
This objective defines a generic paradigm for generative initialization in VQE, guiding the model to prioritize parameter regions that yield low energy across diverse problem instances.
However, optimizing this objective is non-trivial in practice. On real quantum hardware or in black-box evaluation scenarios, we can only observe the scalar measurement statistics: even though the gate sequence we submit is known, the intermediate quantum states --- and thus their derivatives required by automatic differentiation~\cite{baydin2018automaticdiff} --- remain inaccessible. Although quantum gradients with respect to the variational parameters can be estimated using the parameter-shift rule, as previously discussed, this incurs significantly higher computational cost. To address these problems, Flow-VQE uses implicit strategies that transform gradient-free energy feedback into a trainable surrogate objective for model optimization.


\subsection{Policy-gradient optimization}

One solution is to optimize the flow model using an unbiased Monte Carlo estimator~\cite{kleijnen1996optimization}, also known as the REINFORCE algorithm~\cite{williams1992simple}:
\begin{equation} \begin{aligned}
    \nabla_{\boldsymbol{\tau}} & \mathbb{E}_{\boldsymbol{\theta} \sim p_{\boldsymbol{\tau}}(\boldsymbol{\theta} | \boldsymbol{\gamma})}   \langle \psi(\boldsymbol{\theta}) | \hat H(\boldsymbol{\gamma}) | \psi(\boldsymbol{\theta}) \rangle  
    \\& \approx \frac{1}{N}\sum_{i=1}^N \langle \psi(\boldsymbol{\theta}_i) | \hat H(\boldsymbol{\gamma}) | \psi(\boldsymbol{\theta}_i) \rangle  \nabla_{\boldsymbol{\tau}} \log p_{\boldsymbol{\tau}}(\boldsymbol{\theta}_i | \boldsymbol{\gamma}). 
\end{aligned} 
    \label{estimator}
\end{equation}
From a policy-learning perspective, $p_{\boldsymbol{\tau}}(\boldsymbol{\theta} | \boldsymbol{\gamma})$ acts as a conditional policy for generating variational parameters, with the energy measurement serving as a negative reward signal. When weighted by the (negative) energy, the gradient term $\nabla_{\boldsymbol{\tau}} \log p_{\boldsymbol{\tau}}(\boldsymbol{\theta}_i | \boldsymbol{\gamma})$ ensures that parameter vectors yielding lower energies exert a greater influence on the model update, thereby biasing the policy toward favorable regions of the variational parameter space. Importantly, the gradient is taken with respect to the log-likelihood of the sampled parameters, rather than directly on the energy itself, enabling principled optimization of generative models in non-differentiable quantum settings.


\subsubsection{Limitations of (online) policy gradient}

Despite the theoretical validity of policy-gradient methods, their application in Flow-VQE faces two practical obstacles specific to quantum chemistry: (i) weak optimization signals, and (ii) high variance in gradient estimates.
Quantum chemistry tasks typically require exceptionally high energy precision to make meaningful predictions~\cite{pople1999nobel, helgakermolecular}. For instance, a mean-field Hartree--Fock (HF) solution typically recovers more than\SI{99.5}{\percent} of the total electronic energy for small closed-shell molecules, while the remaining correlation energy is critical for accurate property predictions~\cite{helgakermolecular,martin2022electron}. However, such subtle energy differences translate to extremely weak learning signals in the high-dimensional parameter landscape, resulting in nearly flat reward landscapes around optimal parameter regions.
Although variance-reduction techniques such as baseline subtraction are common in reinforcement learning~\cite{sutton1998reinforcement}, the effectiveness of these baselines relies on stable empirical estimates, which become statistically unreliable under the small batch sizes imposed by quantum resource limitations.

In addition, the Monte Carlo estimator in \eqref{estimator} requires extensive samples to achieve acceptable variance bounds. The online nature of the estimator --- where updates are based solely on samples drawn in the current iteration --- amplifies sampling inefficiency during early training, when most samples lie far from the optimum. Taken together, these factors result in a vanishing signal-to-noise ratio in gradient estimation, severely impeding learning dynamics and rendering online policy-gradient methods inefficient in low-signal, resource-constrained quantum environments.


\subsection{Preference-based optimization}
\label{pbo}

Drawing inspiration from recent advances in preference fine-tuning for large language models~\cite{rafailov2023direct, xu2024contrastive}, we propose a preference-based optimization approach for Flow-VQE to mitigate the limitations of the policy-gradient method. Rather than relying on high-variance energy-weight signals, our approach leverages direct performance preferences among sampled parameters to construct a more informative supervision signal. 

Specifically, we define a preference relation:
$\boldsymbol{\theta}_i \succ \boldsymbol{\theta}_j$ if $E(\boldsymbol{\theta}_i) < E(\boldsymbol{\theta}_j)$, where $E(\boldsymbol{\theta})$ is the energy expectation value measured for a given sample $\boldsymbol{\theta}$. We maintain and dynamically update an `elite memory buffer' $\mathcal{B} = \{(\boldsymbol{\theta}^w_k, \boldsymbol{\gamma}_k)\}_{w=1}^M$, which stores the top $M$ samples under the energy-based preference ordering for each molecular structure $k$.
To optimize the Flow-VQE distribution $p_{\boldsymbol{\tau}}(\boldsymbol{\theta}|\boldsymbol{\gamma})$, we perform MLE over the samples from buffer $\mathcal{B}$:
\begin{equation} 
\mathcal{L}(\boldsymbol{\tau}) = - \mathbb{E}_{(\boldsymbol{\theta}^w_k, \boldsymbol{\gamma}_k) \sim \mathcal{B}} \mleft[\log p_{\boldsymbol{\tau}}(\boldsymbol{\theta}^w_k | \boldsymbol{\gamma}_k)\mright] ,
\label{eq:preference_loss}
\end{equation}
where the log-likelihood is explicitly calculated by \eqref{eq:logp}. 

We clarify that the distinction between standard MLE [\eqref{eq:nll}] and our approach lies in the source of training data: while standard generative modeling focuses on samples from a fixed target data set, our method performs MLE on a self-sampled data set whose quality is progressively enhanced through dynamic selection based on the preference criterion. Notably, both the objectives in \eqref{estimator} and \eqref{eq:preference_loss} are consistent with \eqref{eq:8}, as each encourages the model to generate higher-quality samples, albeit through distinct supervision mechanisms.

Preference optimization decouples model updates from immediate online sampling, reducing distributional drift and gradient variance by focusing exclusively on high-quality samples. This avoids the instability often observed in online policy methods, when low-quality samples dominate under limited sampling budgets. Furthermore, by replacing noisy scalar rewards with binary comparisons; the preference-based objective amplifies learning signals even in low-energy-differential regimes. As a result, the model can improve reliably with small exploratory batches, substantially enhancing sample efficiency in quantum-constrained settings.

The pseudocode for Flow-VQE training via preference optimization is presented in Algorithm~\ref{algo:1}. We note that, compared to conventional VQE training, Flow-VQE introduces some additional hyperparameters, such as buffer size, batch size, and architectural parameters of the flow model. In practice, this increases implementation complexity and may require more empirical tuning across different systems. On the other hand, it opens up opportunities for further improvements through automated hyperparameter tuning~\cite{yu2020hyper} and model-architecture refinement.

\begin{algorithm} 
\caption{Training Flow-VQE based on preference optimization.}
\label{algo:1}
\begin{algorithmic}[1]
\REQUIRE Molecular conditions $\{\boldsymbol \gamma_k\}_{k=1}^{K}$, Hamiltonians $\{\hat H(\boldsymbol\gamma_k)\}_{k=1}^{K}$, flow model $p_{\boldsymbol\tau}(\boldsymbol\theta | \boldsymbol\gamma)$, training epochs $T$, batch size $B$, buffer size $M$
\STATE Initialize flow parameters $\boldsymbol\tau$, memory buffers $\mathcal{B}=\{D_k = \emptyset\}_{k=1}^{K}$
\FOR{epoch = 1 to $T$}
    \FOR{each $\boldsymbol \gamma_k$ in $\{\boldsymbol \gamma_k\}_{k=1}^{K}$}
        \STATE Sample parameters from the flow model: 
        $\{(\boldsymbol \theta_i^{\mathrm{new}}, \log p_{\boldsymbol \tau} (\boldsymbol \theta_i^{\mathrm{new}}|\boldsymbol \gamma_k))\}_{i=1}^{B} \sim p_{\boldsymbol \tau}(\boldsymbol \theta|\boldsymbol \gamma_k)$
        \FOR{$i = 1$ to $B$}
       \STATE \makebox[\linewidth][l]{Evaluate energy: $E_i^{\mathrm{new}} \leftarrow \langle\psi(\boldsymbol \theta_i^{\mathrm{new}})|\hat H(\boldsymbol \gamma_k)|\psi(\boldsymbol \theta_i^{\mathrm{new}})\rangle$}
       
       \STATE Add to memory buffer:
            $D_k \leftarrow D_k \cup \{(\boldsymbol \theta_i^{\mathrm{new}}, E_i^{\mathrm{new}})\}$
        \ENDFOR
        \STATE Sort $D_k$ by ascending energy
        \IF{$|D_k| > M$}
        \STATE $D_k \leftarrow \{(\boldsymbol \theta_{k}^w, E_{k}^w) \in D_k \mid w \in \{1, \dots, M\}\}$ \COMMENT{Keep top $M$}
        \ENDIF
    \ENDFOR
    \STATE Preference optimization:
    \STATE $\boldsymbol \theta^W \leftarrow \emptyset$
    \FOR{each $D_k$, corresponding to $\boldsymbol \gamma_k$}
        \STATE $\boldsymbol \theta^W \leftarrow \boldsymbol \theta^W \cup \{ (\boldsymbol \theta_{k}^w, \boldsymbol \gamma_k) \mid (\boldsymbol \theta_{k}^w, E_{k}^w) \in D_k\}$  
    \ENDFOR 
    \STATE Compute loss: $\mathcal{L} \leftarrow -\mathbb{E}_{(\boldsymbol \theta_k^w, \boldsymbol \gamma_k) \in \boldsymbol \theta^W}[\log p_{\boldsymbol \tau}(\boldsymbol \theta_k^w| \boldsymbol \gamma_k)]$
    \STATE Update: $\boldsymbol \tau \leftarrow \boldsymbol \tau - \eta\nabla_{\boldsymbol \tau}\mathcal{L}$ 
\ENDFOR
\RETURN $p_{\boldsymbol\tau}(\boldsymbol\theta|\boldsymbol\gamma)$
\end{algorithmic}
\end{algorithm}


\section{Numerical simulations}  
\label{sec:numerical}

In this section, we investigate the performance of Flow-VQE in a dual role: as a standalone optimizer for direct energy minimization, and as a warm-start parameter generator providing high-quality initializations for conventional VQE routines. Our simulations span multiple molecular systems with varying geometric configurations: a linear hydrogen chain (H$_4$), water (H$_2$O), ammonia (NH$_3$) and benzene (C$_6$H$_6$). 
For each system, we explore distinct conformational changes:  simultaneous stretching of neighboring H--H distances in H$_4$; symmetric stretching of both O--H bonds in H$_2$O; nitrogen pyramidal inversion in NH$_3$ while maintaining the three hydrogen atoms fixed in a plane; and stretching of a single C--H bond in C$_6$H$_6$. The corresponding atomic coordinates for all molecules are provided in \appref{app_coordinate}.


\subsection{Simulation setup}


\subsubsection{Electronic structure modeling}

We implement all computational procedures using PennyLane~\cite{bergholm2018pennylane} with OpenFermion plugins~\cite{mcclean2020openfermion}. We employ the minimal STO-3G basis set for H$_2$O, NH$_3$, and C$_6$H$_6$, and the cc-pVDZ basis set for H$_4$ to construct the restricted HF orbitals. For active-space selections, we perform them to manage computational complexity while preserving essential electronic structure features: (4e, 4o) for H$_4$, (6e, 5o) for H$_2$O, (6e, 6o) for NH$_3$, and (6e, 6o) for C$_6$H$_6$.


\subsubsection{Quantum computation aspect}

The fermionic Hamiltonians are mapped to qubit Hamiltonians using the Jordan--Wigner transformation, resulting in the following qubit requirements: 5 qubits for H$_4$ (after applying $\mathbb{Z}_2$ symmetry tapering~\cite{bravyi2017tapering}), 10 qubits for H$_2$O, and 12 qubits for both NH$_3$ and C$_6$H$_6$. 
For the PQC architectures, we employ a 10-layer hardware-efficient $R_Y$-linear ansatz~\cite{kandala2017hardware} for H$_4$ and a Givens-based singles and doubles (GSD) ansatz~\cite{arrazola2022universal,xia2020qubit} for H$_2$O, NH$_3$, and C$_6$H$_6$. 
Corresponding to these ansätze, the number of optimizable variational parameters is $55$ for H$_4$, $54$ for H$_2$O, and $117$ for both NH$_3$ and C$_6$H$_6$. Further details of the ansatz structures are provided in \appref{app_ansatz}.


\subsubsection{Flow-model implementation}

In our implementation of Flow-VQE, we employ Gaussianization flows~\cite{meng2020gaussianization} as the backbone NF architecture. Gaussianization flows support efficient likelihood evaluation for fast training, enable rapid sampling, exhibit greater robustness to data transformations, and generalize more effectively on small datasets than other mainstream flow models~\cite{meng2020gaussianization}. Each element-wise transformation in the flow is parameterized by a multi-layer perceptron comprising three hidden layers with 256 units each and exponential linear unit (ELU) activation functions~\cite{clevert2015fast}. Each flow layer employs a mixture of logistic distributions with 32 components. A detailed description of Gaussianization flows is provided in \appref{app_flow}.
In addition, the base distribution uses a multivariate normal $\mathcal{N}(\boldsymbol{\mu} = \boldsymbol{0}, \boldsymbol{\Sigma} = 0.01 \cdot \boldsymbol{I})$, ensuring that initial samples remain close to the HF reference point, since the zero vector corresponds exactly to HF initialization under our ansätze.

To guide conditional generation, we construct the context vector $\boldsymbol{\gamma}$ for each molecular geometry by passing its Hamiltonian coefficients --- represented in a fixed-order Pauli string basis --- through a simple linear embedding layer. This design ensures a consistent context ordering across different configurations of the same molecule, facilitating generalization across geometric variations.

We train the flow model using the Adam optimizer~\cite{kingma2014Adam}, with both the learning rate $\eta$ and weight decay set to 0.0001; these values are empirically chosen from small-scale simulations. To enhance exploration and reduce overfitting under limited sample regimes, Gaussian noise with zero mean and a variance of 0.001 is added to the winning parameters before evaluating the sample likelihood during Flow-VQE training. In the preference-based optimization setting, all experiments use a batch size of $B = 2$ per sampling and maintain a buffer size of $M = 2$ to retain at most two winning samples per molecular configuration. We deliberately adopt such a small-scale setup to demonstrate the sample efficiency of our method in low-budget scenarios.


\subsection{Experimental design}

We evaluate the Flow-VQE framework under two training regimes tailored to different experimental objectives: Flow-VQE-S, which is trained on a single molecular configuration for direct optimization, and Flow-VQE-M, which is trained on multiple configurations for generative warm starts. We use seven flow layers for H$_4$ and ten layers for H$_2$O, NH$_3$, and C$_6$H$_6$ in Flow-VQE-S. For Flow-VQE-M, we employ 20 flow layers for each molecule. 


\subsubsection{Performance metric}

We adopt the number of quantum circuit evaluations, independent of measurement shot counts, as the primary performance metric for comparing optimization algorithms. This choice is justified by the observation that changes in molecular geometry typically only modify the coefficients in the Hamiltonian, while the operator terms remain unchanged~\cite{nakaji2024generative} --- assuming that the self-consistent field procedure converges without numerical artifacts. Therefore, the statistical requirements for observable measurements remain comparable across different molecular configurations, enabling a consistent evaluation protocol. In contrast, reporting only the number of optimization iterations can be misleading, as different algorithms require varying numbers of quantum circuit executions per iteration to determine parameter updates.


\subsubsection{Single molecule optimization}

We evaluate five optimization algorithms on H$_2$O and H$_2$ molecular systems: gradient descent (GD), quantum natural-gradient simultaneous perturbation stochastic approximation (QNSPSA)~\cite{gacon2021simultaneous}, Adam, and the two variants of our Flow-VQE method (Flow-VQE-S and Flow-VQE-M). While Flow-VQE-M is primarily designed as a parameter generator for warm-starting downstream tasks, it inherently performs multi-objective joint optimization. Accordingly, we include its convergence behavior in our comparative assessment. 

The total number of circuit evaluations required by each method is computed as follows: $2dN_{\mathrm{epoch}}$ for GD and Adam, where $d$ is the variational parameter dimension and $N_{\mathrm{epoch}}$ is the total number of optimization iterations; $6N_{\mathrm{epoch}}$ for QNSPSA, which is a gradient-free algorithm with constant complexity; and $BN_{\mathrm{epoch}}$ for Flow-VQE-S and Flow-VQE-M, with a batch size of $B = 2$. 


\subsubsection{Generative warm start}

We evaluate the generative capability of Flow-VQE-M by analyzing its performance in producing approximate variational parameters for the ansätze along the PESs of H\textsubscript{2}O and H\textsubscript{4} molecular systems. 
For experimental evaluation, we train Flow-VQE-M on six molecular geometries of H\textsubscript{2}O and eight of H\textsubscript{4}, respectively. After 5,000 training epochs (equivalent to 10,000 circuit evaluations per molecule), we evaluate performance by uniformly selecting 50 molecular structures across the bond-length domain. For each structure, we sample 16 parameter vectors from the flow model and assess both the minimum and mean energies as primary metrics. 
We further select high-error samples from H\textsubscript{2}O and H\textsubscript{4} for post-training, aiming to explore the potential to rapidly drive the ansätze toward their expressivity limits under different optimizer settings.


\subsubsection{Estimating cost advantage}

Given the probabilistic nature of Flow-VQE’s inference mechanism, deriving a general analytical expression for cost evaluation is infeasible and inherently dependent on implementation details. To provide a representative estimate, we conduct empirical case studies on the NH$_3$ and C$_6$H$_6$ molecules under lightweight training, where both the number of iterations and training configurations are deliberately limited to avoid generating overly high-quality parameters too early and to retain room for warm-started optimization. 

Specifically, we train Flow-VQE-M on four configurations for each molecule. For NH$_3$, the training points correspond to nitrogen displacements $\{-0.5, -0.17, 0.17, 0.5\}~\mathrm{\AA}$ along the inversion path, where \SI{0}{\angstrom} denotes the geometric center of the hydrogen plane; each point involves 3,000 circuit evaluations (12,000 in total). For C$_6$H$_6$, the configurations involve stretching a single C--H bond relative to its equilibrium length (\SI{1.084}{\angstrom}) by $\{-0.3, 0.0, 0.3, 0.6\}~\mathrm{\AA}$, with 6,000 evaluations per point (24,000 in total).
Provided that the cumulative cost savings enabled by generative warm starts across downstream tasks outweigh the initial training overhead, the Flow-VQE-M approach yields a net advantage in quantum resource utilization.

Additionally, we benchmark Flow-VQE-M against the conventional PT method. In our PT protocol, we choose the \SI{0}{\angstrom} configuration as the reuse point for each molecule and optimize it via Flow-VQE-S, using 2,000 circuit evaluations for NH$_3$ and 6,000 for C$_6$H$_6$ (sufficient to reach computational accuracy), thus defining the pre-training cost for each system.


\section{Results and discussion}
\label{sec:results}

\subsection{Single-molecule optimization}

\begin{figure}  
\centering
\includegraphics[width=\linewidth]{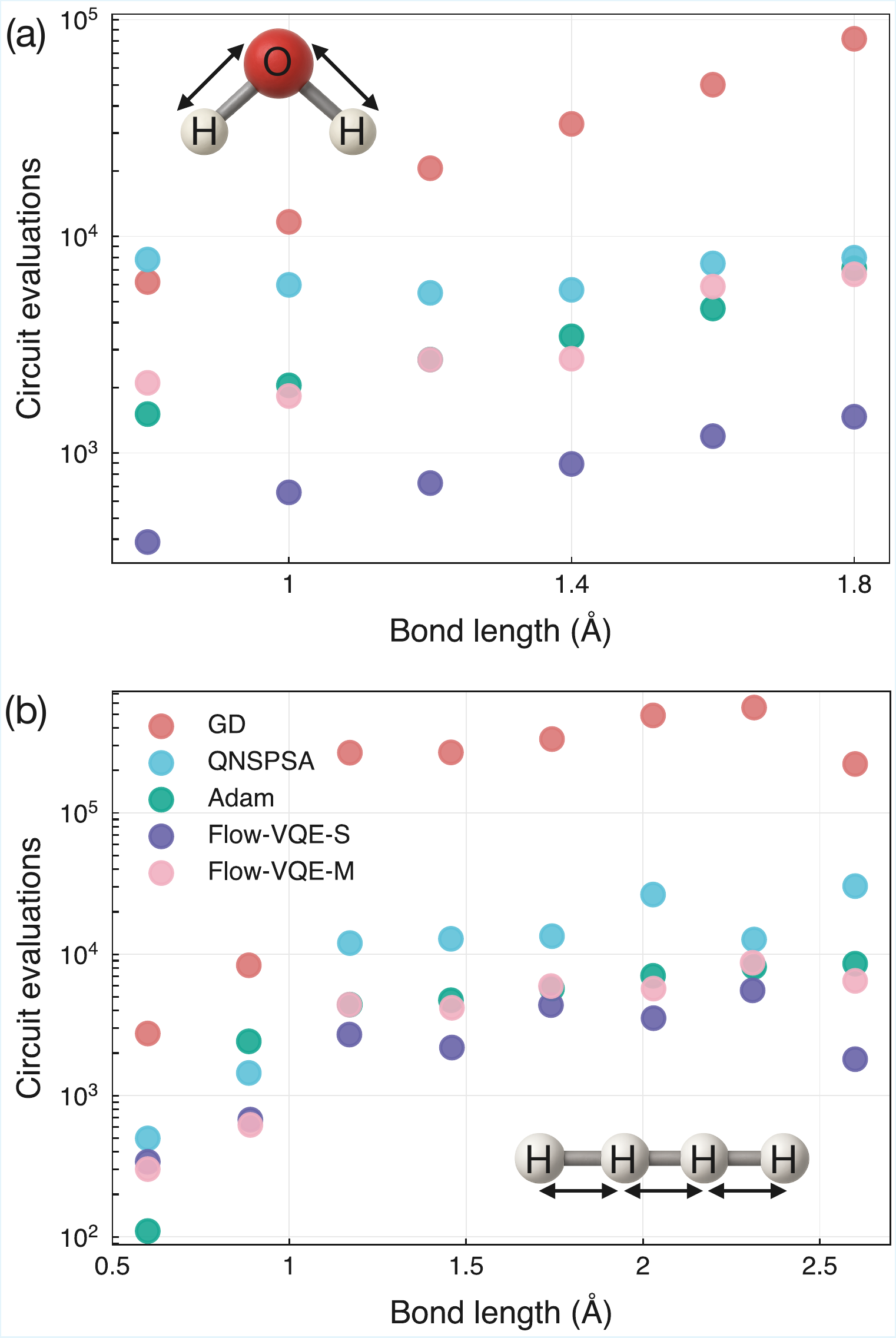}
\caption{Number of quantum circuit evaluations required to achieve computational accuracy for optimizing (a) H$_2$O and (b) H$_4$ using different optimizers. Optimization is performed over six uniformly spaced bond lengths in $[0.8, 1.8]~\mathrm{\AA}$ for H$_2$O and eight lengths in $[0.6, 2.6]~\mathrm{\AA}$ for H$_4$. The arrows in each molecular diagram represent changing bond lengths. All baseline optimizers use a learning rate of $\eta=0.02$.}
  \label{fig:optimizers-comparison}
\end{figure}

Figure~\ref{fig:optimizers-comparison} illustrates the number of circuit evaluations required by each algorithm to achieve convergence within computational accuracy, defined as an error not exceeding $1.6 \times 10^{-3}$ Hartree relative to exact diagonalization at the same level of theory~\cite{lolur2023reference}. Here, the exact diagonalization results are assumed to be known for a benchmarking purpose.

For the H\textsubscript{2}O molecule [\figpanel{fig:optimizers-comparison}{a}], Flow-VQE-S consistently outperforms all baseline optimizers across the test structures, e.g., achieving improvements of one to two orders of magnitude over traditional GD. Compared to the competitive Adam baseline, Flow-VQE-S achieves approximately a two- to five-fold reduction in the number of circuit evaluations. 

Flow-VQE-M also performs comparably to Adam, despite being trained in a more challenging multi-objective setting. While Flow-VQE-M does not match the convergence efficiency of Flow-VQE-S, this is expected: Flow-VQE-M is designed to generalize across multiple structures, rather than optimize a single fixed instance. Thus, it trades off some task-specific performance in favor of broader applicability and transferable warm-start capability.

For the H\textsubscript{4} system [\figpanel{fig:optimizers-comparison}{b}], which spans a broader range of bond lengths (\SI{0.6}{\angstrom} to \SI{2.8}{\angstrom}), Flow-VQE-S generally outperforms the baseline optimizers across most bond lengths. The only exception arises at \SI{0.6}{\angstrom}, where the optimization target is close to the initial state, allowing Adam to converge quickly. While Flow-VQE-S still retains an overall advantage, its margin over Adam is reduced --- an effect that may be attributed to the more rugged energy landscapes introduced by the hardware-efficient ansatz, whose trainability remains a significant challenge. Furthermore, the performance gap between Flow-VQE-M and Flow-VQE-S narrows in this setting, implying that broader distributional exploration may contribute to avoiding certain optimization traps.


\subsection{Generative warm start}


\subsubsection{Generating potential-energy surfaces}

\begin{figure*} 
\centering
\includegraphics[width=\linewidth]{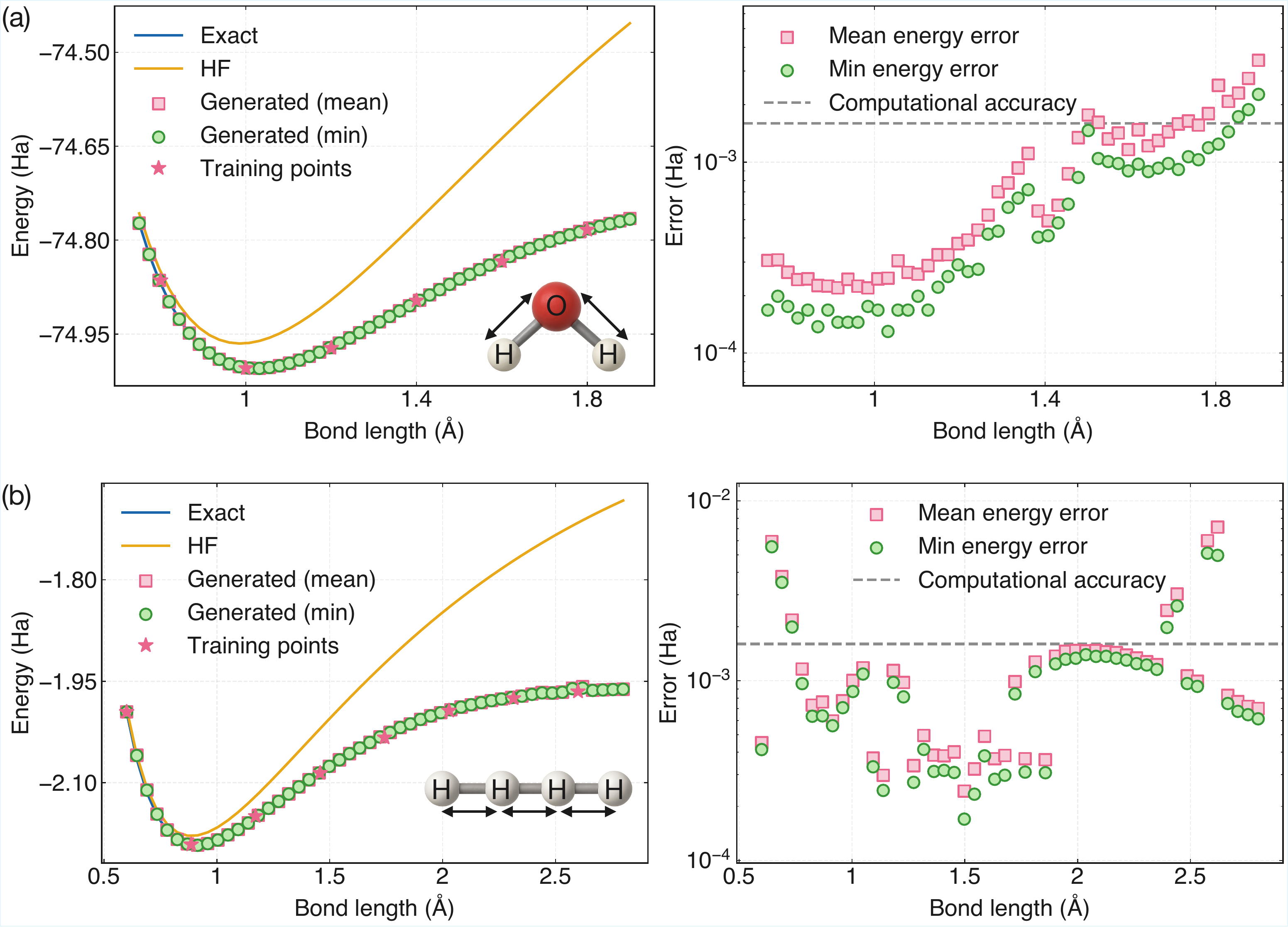}
\caption{Potential-energy surfaces (left) and corresponding errors (right) for (a) H$_2$O and (b) H$_4$ evaluated using parameters generated by Flow-VQE-M. Training points are consistent with those used in \figref{fig:optimizers-comparison}. Generation is performed on 50 uniformly spaced test points in the range $[0.75, 1.9]~\mathrm{\AA}$ for H$_2$O and $[0.6, 2.8]~\mathrm{\AA}$ for H$_4$. In the left panels, the generated points (markers) coincide so closely with the exact curve (blue) that they mostly obscure the latter.}
    \label{fig:PES}
\end{figure*}

The results presented in \figref{fig:PES}, which include the PESs of the representative systems H$_2$O and H$_4$, substantiate the high quality of parameters generated by Flow-VQE-M. When evaluated in the corresponding ansätze, these generated parameters yield energy expectation values closely align with the exact solutions, significantly outperforming HF baselines. Overall, energy errors increase in stretched bond-length regions due to strong correlation effects~\cite{Sheka_2015} and the limited expressivity of the employed ansätze. Nevertheless, Flow-VQE-M maintains high-quality results, demonstrating robust performance even under these challenging conditions.
Notably, most generated points already achieve computational accuracy, leading to high quality PESs. These points can therefore serve as strong initializations for VQE, facilitating rapid convergence to the desired precision.  

\subsubsection{Warm-start post-training}

\begin{table}
    \centering
    \renewcommand{\arraystretch}{1.2}
    \begin{tabular}{c|c|cc|cc}
    \hline\hline
  \multirow{2}{*}{Molecule}   &  \multirow{2}{*}{$\eta$}  & \multicolumn{2}{c|}{$N_{\mathrm{c.a.}}$} & \multicolumn{2}{c}{$\Delta E_{\mathrm{min}}$} \\
    & & FVM & HF & FVM & HF \\
    \hline
    \multirow{3}{*}{H$_2$O} & 0.02 & 324 & 8748 & $5.932\times10^{-4}$ & $5.932\times10^{-4}$ \\
    & 0.005 & 432 & 30672 & $5.935\times10^{-4}$ & $6.000\times10^{-4}$ \\
    & 0.001 & 1944 & 102276 & $5.941\times10^{-4}$ & $1.033\times10^{-3}$ \\
    \hline
    \multirow{3}{*}{H$_4$} & 0.02 & 770 & 8910 & $5.205\times10^{-8}$ & $2.489\times10^{-7}$ \\
    & 0.005 & 440 & 16720 & $1.319\times10^{-9}$ & $3.202\times10^{-9}$ \\
    & 0.001 & 1980 & 72270 & $1.053\times10^{-10}$ & $2.579\times10^{-4}$ \\
    \hline\hline
    \end{tabular}
    \caption{Post-training comparison of Flow-VQE-M (FVM) and HF initializations for H$_2$O at a bond length of $\SI{1.90}{\angstrom}$ and H$_4$ at $\SI{2.58}{\angstrom}$ using different learning rates ($\eta$). After obtaining initial parameter sets from either FVM or HF, we perform standard VQE by running 1000 iterations of the Adam optimizer. The performance metrics reported are the number of circuit evaluations required to reach computational accuracy ($N_{\mathrm{c.a.}}$) and the minimum energy error achieved over iterations ($\Delta E_{\mathrm{min}}$).}
    \label{tab:post_fine_tuning}
\end{table}

Table~\ref{tab:post_fine_tuning} presents a comparison of post-training performance between Flow-VQE-M and HF initialization at different learning rates on H\textsubscript{2}O and H\textsubscript{4}. Across all settings, Flow-VQE-M consistently demonstrates faster convergence and lower energy errors compared to HF initialization. In practice, learning rates $\eta$ are typically chosen between 0.01 and 0.1; for our benchmarks we set $\eta=0.02$. At this rate, Flow-VQE-M reduces the number of circuit evaluations required to reach computational accuracy by over 27-fold for H\textsubscript{2}O and more than 11-fold for H\textsubscript{4}. Moreover, this advantage grows even more pronounced at smaller $\eta$.  For instance, at $\eta=0.001$, Flow-VQE-M cuts the required evaluations by more than 50-fold for H\textsubscript{2}O and over 36-fold for H\textsubscript{4}, while achieving significantly lower minimum energy errors. Complete optimization trajectories corresponding to the results in Table~\ref{tab:post_fine_tuning} are further provided in \appref{app_post_training}.
These results underscore the effectiveness of Flow-VQE-M as a robust warm-start strategy that substantially enhances sample efficiency and convergence precision in the post-training stage, even under conservative optimization settings.


\subsection{Estimate of cost advantage}

To provide a rough quantitative assessment, we use the number of circuit evaluations required to reach a target computational accuracy as the cost metric. As shown earlier in \figref{fig:optimizers-comparison}, Flow-VQE-M incurs a number of evaluations during training comparable to that of standard VQE with Adam. This suggests that, when training points are also required for a task, Flow-VQE-M offers a clear quantum resource advantage due to its additional generative capability. In what follows, we focus exclusively on the cost evaluation for unseen test configurations.

\begin{figure*} 
    \centering  
    \includegraphics[width=\linewidth]{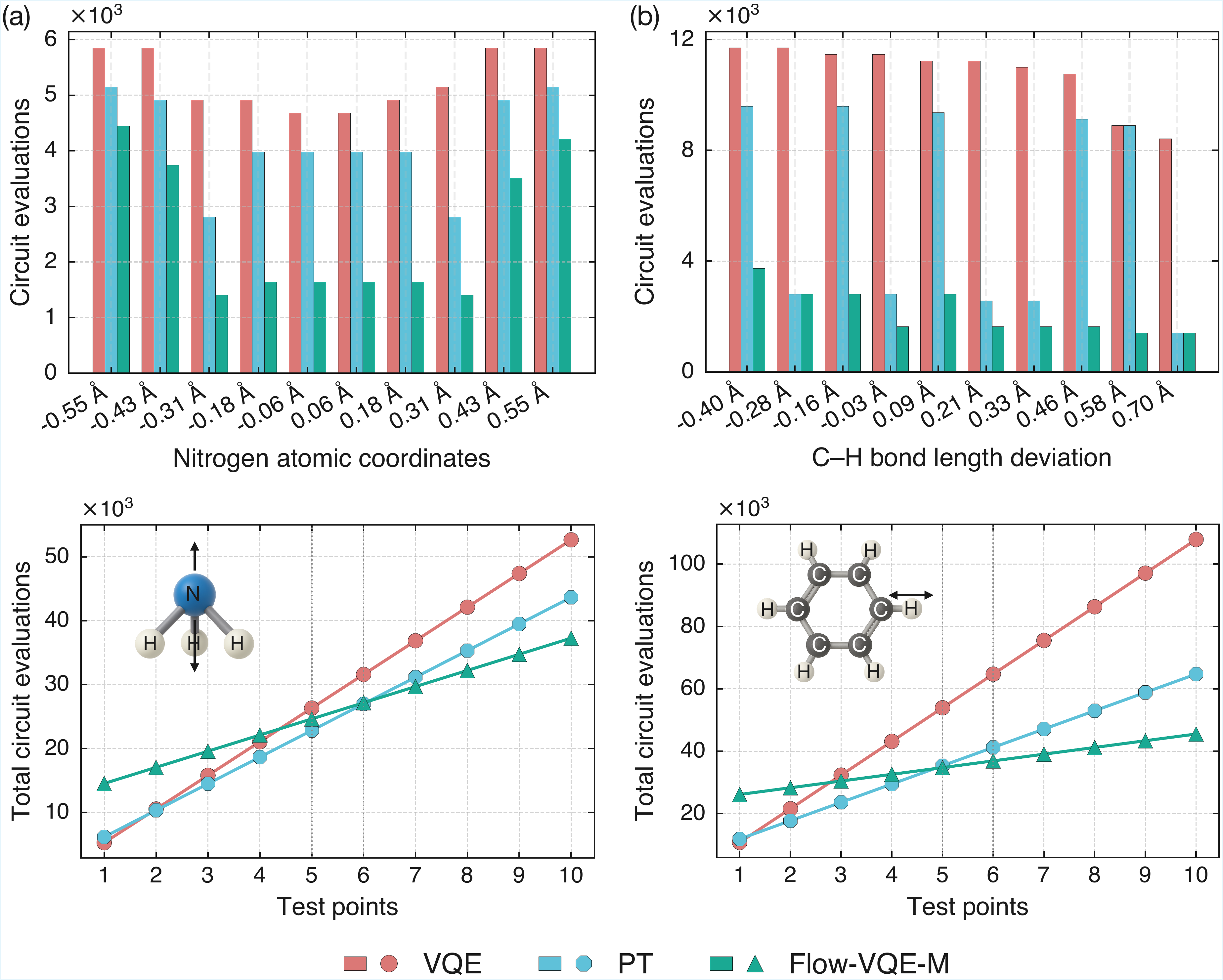}    
    \caption{Circuit evaluation costs for achieving computational accuracy in (a) NH$_3$ and (b) C$_6$H$_6$, comparing HF, PT, and Flow-VQE-M initializations. The upper panels show circuit evaluation counts across different molecular geometries. The lower panels present approximated average scaling trends as the number of test points increases; the intercept indicates the one-time pre-training cost for each strategy, and the slope reflects the average post-training cost per additional test point. The arrows in each molecular diagram represent modes of atomic displacement. Optimization is performed using the Adam optimizer with a learning rate of $\eta = 0.02$. Note that evaluation counts are recorded as integer multiples of the number of variational parameters; hence, circuit evaluations with similar initialization contributions differ by less than a full iteration, resulting in bars of identical height.}
    \label{fig:post_training_cost}
\end{figure*}

Let $C_{\mathrm{pre}}$ denote the total pre-training cost, and $\bar C_\mathrm{post}$ the average post-training cost, per test point. Although individual VQE procedures require significantly different optimization efforts, we approximate the cost growth using average values for simplicity. Accordingly, the total cost of using a typical warm-start method (such as Flow-VQE or PT) for $n_{\mathrm{test}}$ test points is $C_\mathrm{total}=C_\mathrm{pre}+ \bar C_\mathrm{post} \cdot n_{\mathrm{test}}$. In contrast, standard VQE incurs a cost of approximately $\bar C_{\mathrm{VQE}} \cdot n_{\mathrm{test}}$. Given that $\bar C_\mathrm{post} < \bar C_{\mathrm{VQE}}$, the gentler cost scaling of warm-start methods highlights their growing advantage as $n_{\mathrm{test}}$ increases.

Figure~\ref{fig:post_training_cost} compares the training costs using standard VQE, PT, and Flow-VQE-M warm-start strategies for NH$_3$ and C$_6$H$_6$, along with corresponding scaling estimates.
For NH$_3$, standard VQE requires an average of $\bar{C}_{\mathrm{VQE}} = 5,265$ circuit evaluations per test point, reduced to $\bar{C}_{\mathrm{post}} = 2,527$ by Flow-VQE-M and $\bar{C}_{\mathrm{post}} = 4,165$ by PT. With pre-training costs of $C_{\mathrm{pre}} = 12,000$ for Flow-VQE-M and $C_{\mathrm{pre}} =2,000$ for PT, Flow-VQE-M achieves a net cost advantage over standard VQE beyond five test points, and over PT beyond six points.
For C$_6$H$_6$, the respective values are $\bar{C}_{\mathrm{VQE}} = 10,787$, $\bar{C}_{\mathrm{post}} = 2,153$ (Flow-VQE-M), and $\bar{C}_{\mathrm{post}} = 5,873$ (PT), with corresponding pre-training costs of $24,000$ and $6,000$. Here, Flow-VQE-M outperforms standard VQE after three test points and PT after five points. These estimates are instance-dependent and not intended as universal benchmarks, but they illustrate the practical advantages of Flow-VQE-M in scenarios requiring repeated evaluations across chemical space. 

We attribute these savings to Flow-VQE-M’s ability to embed problems into a latent space where similarity captures the closeness of the optimal variational parameters, a shift from conventional PT. Unlike PT, which relies on heuristic reuse based on geometrical proximity, Flow-VQE-M conditions its parameter distribution on rich, task-specific embeddings spanning diverse molecular configurations. While these embeddings are not explicitly engineered to encode structural or electronic features, the model benefits from exposure to a broad training distribution, allowing it to capture latent task similarities. Notably, advances in molecular descriptors~\cite{wang2021quantum, von2020exploring} provide a fertile ground for designing more informative embeddings. Nevertheless, even in its current form, Flow-VQE-M is capable of generalizing beyond geometric similarity. This is exemplified in the case of C$_6$H$_6$, where reused parameters from the \SI{0}{\angstrom} configuration yields unexpectedly fast convergence for distant configurations, suggesting the presence of deeper task correlations not captured by geometry alone. In contrast to the manual identification of transferable patterns required by PT, Flow-VQE-M learns how contextual embeddings modulate the parameter distribution, enabling more consistent and scalable transfer across chemically diverse settings.


\section{Limitations and future work}
\label{sec:future}

Flow-VQE, as a black-box optimizer for VQE, may be less precise than gradient-based methods in smooth, differentiable energy landscapes, where exact gradients enable fine exploitation of local curvature. However, its gradient-free nature may offer improved robustness for noisy VQE --- a hypothesis that remains to be systematically validated. 

In addition, the preference-based optimization in Flow-VQE tends to concentrate probability density over time, which can reduce sample diversity and limit exploratory behavior. Mitigating this limitation requires further investigation into entropy-regularized objectives or diversity-promoting preference selection strategies~\cite{lanchantin2025diverse, maus2022discovering}.

Once sampling diversity is better preserved, Flow-VQE can become a natural front-end to quantum subspace methods~\cite{motta2024subspace, robledo2025chemistry}. A trained flow model can generate many parameter sets in constant time, each yielding a trial state with high ground-state overlap. These non-orthogonal yet independent states can span a compact subspace, thereby lowering the effective dimensionality of the subsequent eigenvalue problem.

While the numerical experiments reported here extend only to 12-qubit active spaces and 117 variational parameters, several features of Flow-VQE promise broad quantum-resource savings, even when scaling to larger molecules. First, each optimization epoch engages the quantum processor only $B$ times --- each engagement itself comprising the required number of shots for sufficient measurement statistics.  Although we use $B=2$ throughout this study, $B$ (and the shot budget) can be dynamically scheduled. During the exploratory phase one may set a small size of $B$ to survey the landscape cheaply, and then raise $B$ once the parameters approach the ground-state basin, thereby sharpening energy estimates without expending resources on low-overlap regions. Because this schedule is independent of the variational dimension $d$, it bypasses the quartic growth in the qubit counts that typically characterizes chemically motivated ansätze \cite{anand2022quantum} and the even steeper scaling of many hardware-efficient ansätze \cite{sim2019expressibility}.
On the classical side, the classical overhead in modern normalizing-flow architectures scales linearly in $d$, keeping training and inference practical even as $d$ reaches the tens of thousands \cite{zhai2025nfcapablegenerative}. 

In addition, after training on a set of representative systems, Flow-VQE can warm start to unseen members of the same chemical family; in principle, this capability persists across system sizes, although retaining predictive power requires explicit molecular representation learning \cite{wang2021quantum,von2020exploring} to distill the most relevant features. In particular, when approximate reference data are available --- such as historically calculated inaccurate data --- the flow model can be pre-trained entirely with classical computation on this corpus, and then fine-tuned through its standard interaction with the quantum device. This two-stage regimen shifts the bulk of optimization off-hardware and making larger systems economically accessible. Capitalizing on these gains, our immediate priority is to scale Flow‑VQE to systems exceeding 20 qubits, thereby furnishing a decisive, empirical assessment of its ability to treat larger, more chemically realistic molecules.

We want to emphasize that the most stubborn training obstacles in VQE
(barren plateaus in particular) stem largely from the ansatz structure \cite{larocca2024review,holmes2022connecting}: merely changing optimization methods cannot, by itself, restore meaningful gradients in an intrinsically flat landscape. Genuine progress therefore requires Flow-VQE and ansatz designs to advance together. 
Accordingly, a compelling extension of Flow-VQE lies in moving beyond fixed parameterized templates to directly generate quantum circuits as symbolic gate sequences. This unified “operator space” approach treats all allowed gate primitives and ordering choices as part of a single search domain, enabling end-to-end optimization of circuit structures. By conditioning generative flows on Hamiltonian features, the model can learn systematic mappings from problem instances to circuit architectures that respect symmetries and generalize across different molecules. In this setting, Flow-VQE offers a principled foundation for discrete sequence modeling and quantum architecture search~\cite{martyniuk2024quantum}, where circuit structures and operator orderings are inherently non-differentiable and combinatorial. Prior studies have shown that operator ordering significantly impacts variational expressivity and simulation accuracy~\cite{grimsley2019trotterized, grimsley2019adaptive, campbell2019random, nakaji2024generative}. Leveraging discrete normalizing flows~\cite{tran2019discrete, hoogeboom2021argmax}, Flow-VQE can be extended to explore such combinatorial spaces efficiently, supporting the automated discovery of expressive, transferable, and Hamiltonian-aware quantum circuits.


\section{Conclusion}
\label{sec:conclusion}

We introduce Flow-VQE, which provides a probabilistic framework for recasting variational quantum optimization as a generative modeling task. By leveraging preference-based training, Flow-VQE eliminates the need for quantum gradient estimation and progressively refines its sampling distribution to generate increasingly high-quality variational parameters. Evaluated on representative molecular systems, Flow-VQE enables lower quantum resource cost during optimization than standard methods and supports transferable warm starts by extracting generalizable features from diverse molecular structures. When generating hundreds of variational parameters, the associated classical training remains easily tractable with modern machine learning techniques and does not pose a computational bottleneck. These results position Flow-VQE as a practical and resource-efficient approach for near-term variational quantum simulation. 

Although the present work focuses on VQE, the framework extends naturally to other variational quantum algorithms --- such as the quantum approximate optimization algorithm \cite{farhi2014} --- whenever a well-defined loss function exists. Extending the flow model from parameter space to operator or mixed space further lays a principled foundation for the emerging field of multimodal quantum circuit generation \cite{fürrutter2025synthesis}. 
Overall, this work highlights the promise of integrating generative modeling with quantum optimization to expand the algorithmic design space and advance hybrid quantum-classical computing.


\begin{acknowledgments}

This work was partially supported by the Wallenberg AI, Autonomous Systems and Software Program (WASP) funded by the Knut and Alice Wallenberg Foundation. The work was further supported by a joint project between WASP and the Wallenberg Centre for Quantum Technology (WACQT), funded by the Knut and Alice Wallenberg Foundation. Preliminary results were enabled by resources provided by the National Academic Infrastructure for Supercomputing in Sweden (NAISS) at Alvis (project: NAISS 2025/22-480), partially funded by the Swedish Research Council through grant agreement no.~2022-06725. AFK also acknowledges support from the Swedish Foundation for Strategic Research (grant numbers FFL21-0279 and FUS21-0063) and the Horizon Europe programme HORIZON-CL4-2022-QUANTUM-01-SGA via the project 101113946 OpenSuperQPlus100. SO thanks Morten Kjaergaard (UCPH) for discussions during the early conceptualization of the presented work.

\end{acknowledgments}
 

\appendix

\section{Molecular geometries \label{app_coordinate}}

We here summarize the parameterized geometries used for selected molecular systems in this work. All coordinates are given in Cartesian format (in \AA), as functions of a distortion parameter (typically a bond length or stretching coordinate). 


\subsection{H\textsubscript{2}O (Water)}

The H--O--H bond angle is fixed at $\theta = 104.5^\circ$, and the O--H bond length is varied symmetrically by a scalar distance $d$. The coordinates are
\[
\begin{bmatrix}
\text{O} & 0 & 0 & 0 \\
\text{H} & d \cdot \sin(\theta/2) & 0 & d \cdot \cos(\theta/2) \\
\text{H} & -d \cdot \sin(\theta/2) & 0 & d \cdot \cos(\theta/2)
\end{bmatrix}.
\]


\subsection{H\textsubscript{4} (Linear hydrogen chain)}

The four hydrogen atoms are aligned along the $x$ axis with uniform spacing $d$. The coordinates are
\[
\begin{bmatrix}
\text{H} & 0 & 0 & 0 \\
\text{H} & d & 0 & 0 \\
\text{H} & 2d & 0 & 0 \\
\text{H} & 3d & 0 & 0
\end{bmatrix}.
\]


\subsection{NH\textsubscript{3} (Ammonia)}

The nitrogen atom is placed at $(0, 0, d)$ along the $z$ axis, while the three hydrogen atoms form an equilateral triangle in the $xy$ plane. The coordinates are
\[
\begin{bmatrix}
\text{N} & 0 & 0 & d \\
\text{H} & 1 & 0 & 0 \\
\text{H} & -0.5 & \sqrt{3}/2 & 0 \\
\text{H} & -0.5 & -\sqrt{3}/2 & 0
\end{bmatrix}.
\]


\subsection{C\textsubscript{6}H\textsubscript{6} (Benzene)}

The equilibrium geometry is obtained from the National Institute of Standards and Technology database~\cite{lemmon2018nist}. The coordinates are
\[
\begin{bmatrix}
\text{C1} & \phantom{-}1.3970 & \phantom{-}0 & 0 \\
\text{C2} & \phantom{-}0.6985 & \phantom{-}1.2098 & 0 \\
\text{C3} & -0.6985 & \phantom{-}1.2098 & 0 \\
\text{C4} & -1.3970 & \phantom{-}0 & 0 \\
\text{C5} & -0.6985 & -1.2098 & 0 \\
\text{C6} & \phantom{-}0.6985 & -1.2098 & 0 \\
\text{H1} & 2.4810+d & \phantom{-}0  &  0  \\
\text{H2} & \phantom{-}1.2405 & \phantom{-}2.1486 & 0 \\
\text{H3} & -1.2405 & \phantom{-}2.1486 & 0 \\
\text{H4} & -2.4810 & \phantom{-}0 & 0 \\
\text{H5} & -1.2405 & -2.1486 & 0 \\
\text{H6} & \phantom{-}1.2405 & -2.1486 & 0 \\
\end{bmatrix}
\]
Here, the equilibrium C--H bond length is \SI{1.084}{\angstrom}. The variable $d$ controls the displacement of H1 along the C1--H1 bond direction, allowing both bond compression ($d < 0$) and stretching ($d > 0$), while all other atoms remain fixed.

\section{Ansatz details \label{app_ansatz}}

\subsection{Givens-based singles and doubles ansatz}

\begin{figure*}
  \centering
  \includegraphics[width=\linewidth]{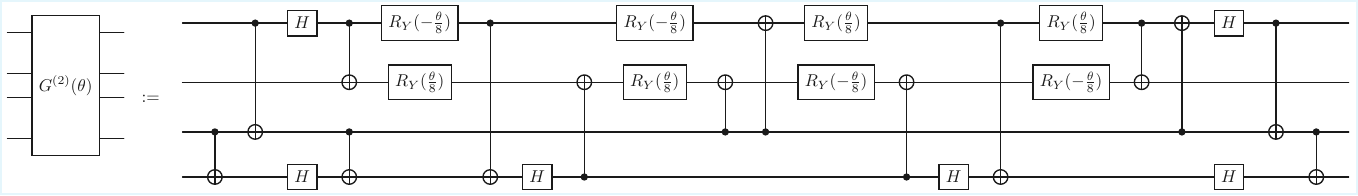}
  \caption{Decomposition of the four-qubit entangling gate $G^{(2)}(\theta)$ into Hadamard ($H$), $R_Y$ rotational gates, and CNOT gates.}
  \label{fig:g2_gate}
\end{figure*}

The Givens-based singles and doubles (GSD) ansatz~\cite{arrazola2022universal,xia2020qubit,zou2025multireference} provides a qubit-native alternative to the conventional unitary coupled-cluster singles and doubles (UCCSD) ansatz~\cite{anand2022quantum}. It constructs the variational wavefunction by sequentially applying particle-conserving exchange gates that encode single and double excitations directly on qubits:
\begin{equation}\begin{aligned}
    |\psi (\boldsymbol{\theta})\rangle &=\mleft(\prod_{(i,j)\in\mathcal{S}}G_{ij}^{(1)}(\theta_{ij})\mright)\\
    &\quad \mleft(\prod_{(k,l,m,n)\in\mathcal{D}}G_{klmn}^{(2)}(\theta_{klmn})\mright)|\psi_0\rangle,
\end{aligned} 
\end{equation}
where $\mathcal{S}$ and $\mathcal{D}$ denote the sets of spin-adapted single and double excitations, respectively. Each gate acts locally and preserves particle number, enabling efficient, constant-depth implementation on quantum hardware. The two-qubit gate $G^{(1)}(\boldsymbol{\theta})$ for single excitations mixes $\ket{01}$ and $\ket{10}$ and can be implemented by 
\begin{equation}
    \begin{aligned}
G^{(1)}(\theta)& =\begin{pmatrix}1&0&0&0\\0&\cos(\theta/2)&-\sin(\theta/2)&0\\0&\sin(\theta/2)&\cos(\theta/2)&0\\0&0&0&1\end{pmatrix} \\
&:= \begin{quantikz} 
         & &\targ{}&\gate{R_Y({\theta}/{2})}&\targ{}&&\\
         & \gate{H}&\ctrl{-1}&\gate{R_Y( {\theta}/{2})}&\ctrl{-1}&\gate{H}&\\ 
    \end{quantikz}.
\end{aligned}
\label{eq6}
\end{equation} 
Double excitations can be implemented using four-qubit gates $G^{(2)}(\theta)$, which coherently mix configurations like $|0011\rangle \leftrightarrow |1100\rangle$. A decomposition choice of $G^{(2)}(\theta)$ into elementary quantum gates is shown in \figref{fig:g2_gate}.

In contrast, the Trotterized UCCSD ansatz requires mapping fermionic excitations to qubits via the Jordan--Wigner transformation~\cite{jordan1928paulische, tranter2018comparison}, where each term introduces long parity strings to preserve anticommutation relations. In the first-order Trotter decomposition, these nonlocal structures lead to a gate complexity upper bounded by $\mathcal{O}(n^5)$ for $n$ spin orbitals~\cite{cao2019quantum, xia2020qubit}. The Givens-based ansatz achieves a reduced gate complexity of $\mathcal{O}(n^4)$~\cite{xia2020qubit}, corresponding to the total number of spin-adapted single and double excitations. This makes it a more hardware-efficient choice for near-term quantum devices. 


\subsection{Hardware-efficient ansatz}

A hardware-efficient ansatz (HEA) is an architecture-driven variational circuit designed to maximize compatibility with the native gate set and connectivity of near-term quantum hardware~\cite{kandala2017hardware}.
It typically consists of alternating layers of parameterized single-qubit rotations $R(\boldsymbol{\theta})$ and entangling gates $U_{\mathrm{ent}}$ arranged in patterns that match the hardware topology. 
The general form of an HEA circuit with $L$ layers is 
\begin{equation}
    |\psi(\boldsymbol{\theta})\rangle = \mleft( \prod_{\ell=1}^{L} R(\boldsymbol{\theta}_\ell) U_{\mathrm{ent}}^{(\ell)}  \mright) R(\boldsymbol{\theta}_0) |\psi_0\rangle.
\end{equation}
In this work, we adopt a hardware-efficient $R_Y$-linear form, where $R(\boldsymbol{\theta}_\ell)=\bigotimes_{i=1}^nR_{Y,i}(\theta_{\ell,i})$ with $n$ qubits, and $U^{(\ell)}_{\mathrm{ent}}$ consists of linear nearest-neighbor CNOT gates.


\subsubsection{Trainability challenges}

While HEAs can become increasingly expressive with depth, it has been shown that sufficiently deep, randomly parameterized circuits can approximate unitary 2-designs~\cite{dankert2009exact} --- ensembles of unitaries that reproduce the second-order statistical moments of the Haar distribution~\cite{holmes2022connecting}. This level of expressivity induces a concentration of measure in the cost landscape, causing barren plateaus~\cite{mcclean2018barren, larocca2024review} --- gradients vanish exponentially with system size. This renders variational training difficult or even infeasible for large circuits. To mitigate this issue, practical implementations often constrain circuit depth, incorporate symmetry-preserving structures, or employ initialization strategies that prevent the circuit from starting in overly entangled or highly randomized regions of Hilbert space~\cite{larocca2024review}. The latter is one of the primary motivations behind warm-start strategies. In addition, we emphasize that chemically inspired ansätze, such as UCCSD, are not immune to barren plateaus, particularly when a large number of excitations are included~\cite{mao2024towards}.


\section{Gaussianization flows \label{app_flow}}

Gaussianization flows (GFs)~\cite{meng2020gaussianization} exhibit dual efficiency for training and sampling, while simultaneously maintaining universal approximation guarantees for continuous distributions.  The model builds upon rotation-based iterative Gaussianization (RBIG)~\cite{laparra2011iterative}, which alternates between marginal Gaussianization and orthogonal transformation steps to transform data toward a standard normal distribution. Gaussianization flows reformulate this process by replacing non-parametric density estimation with trainable mixtures of logistic distributions and substituting fixed rotations with learnable orthogonal transformations parameterized via Householder reflections~\cite{householder1958unitary,tomczak2016improving}.

Formally, GFs construct the bijective mapping by a sequence of alternating layers:
\begin{equation}
f_{\boldsymbol{\tau}}=\Psi_{\boldsymbol{\tau}_K} \circ R_K \circ \cdots \circ \Psi_{\boldsymbol{\tau}_1} \circ R_1 .
\end{equation}
Each marginal Gaussianization layer $\Psi_{\boldsymbol{\tau}_i}(\mathbf x)$ applies the transformation
\begin{equation}
\Psi_{\boldsymbol{\tau}_i}(\mathbf x)= \mleft(\Psi^{(1)}_{\boldsymbol{\tau}_i}( x ^{(1)}), \ldots, \Psi^{(d)}_{\boldsymbol{\tau}_i}(x ^{(d)})\mright)^{\top}, 
\end{equation} 
where each dimension-wise component is defined as 
\begin{equation}
\Psi_{\boldsymbol{\tau}_i}^{(k)}\mleft(x^{(k)}\mright)=\Phi^{-1} \circ F_{\boldsymbol{\tau}_i}^{(k)}\mleft(x^{(k)}\mright), \quad
k=1, \ldots, d.
\end{equation}
Here, $\Phi^{-1}$ denotes the inverse of the Gaussian cumulative density function (CDF), while
\begin{equation}
F_{\boldsymbol{\tau}_i}^{(k)}\mleft(x^{(k)}\mright)=\frac{1}{P} \sum_{j=1}^P \sigma\mleft(\frac{x^{(k)}-\mu_{j}^{(k)}}{h_{ j}^{(k)}}\mright)
\end{equation}
parameterizes a CDF through a mixture of $P$ logistic components with learnable anchor points $\mleft\{\mu_{j}^{(k)}\mright\}_{j=1}^P$ and bandwidths $\mleft\{h_{j}^{(k)}\mright\}_{j=1}^P$, where $\sigma(\cdot)$ represents the logistic sigmoid function. Each orthogonal matrix $R_i$ is efficiently parameterized as a product of Householder transformations~\cite{householder1958unitary, tomczak2016improving}.
The orthogonal transformations maintain invertibility while enabling efficient Jacobian determinant computation, resulting in a model that balances expressivity with computational efficiency in both directions.

\section{Training dynamics \label{app_training}}

\subsection{Flow-VQE training trajectories}

\begin{figure*} 
    \centering
\includegraphics[width=\linewidth]{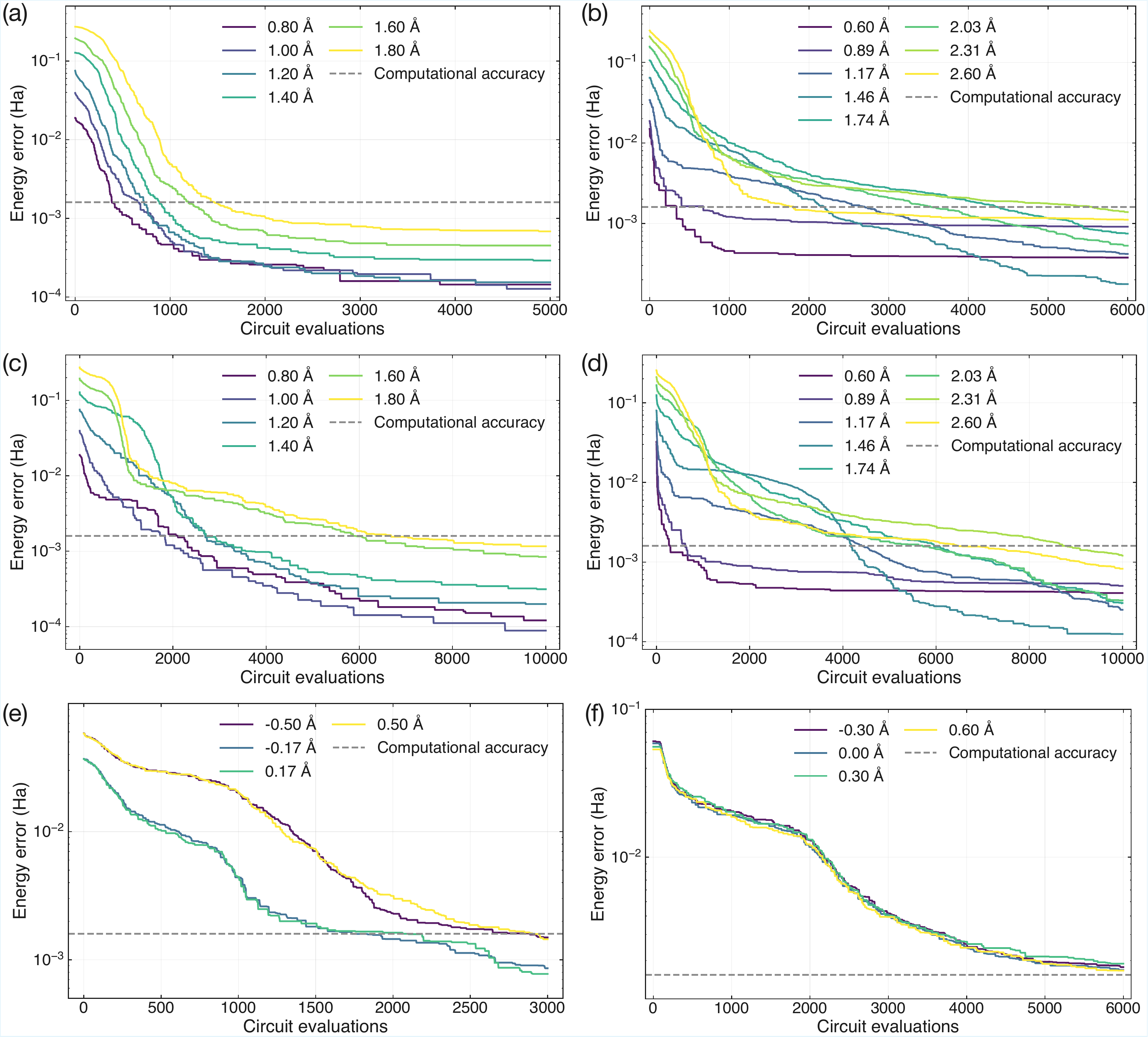}
    \caption{Global minimum energy error monitored throughout Flow-VQE training. (a), (b) Flow-VQE-S for H$_2$O and H$_4$, respectively; (c)--(f) Flow-VQE-M for H$_2$O, H$_4$, NH$_3$, and C$_6$H$_6$, respectively. Flow-VQE-S trains a separate model for each geometry, while Flow-VQE-M employs a single shared model across all geometries for each molecule.}
    \label{fig:min_monitoring}
\end{figure*}

Figure~\ref{fig:min_monitoring} shows the global minimum energy error throughout the training process across all test molecular systems, illustrating the model's convergence behavior and optimization efficiency. The consistent downward trend across diverse bond lengths underscores Flow-VQE’s capability to progressively concentrate probability density in low-energy regions, thereby improving the quality of variational parameters over time. 

However, local plateau regions are evident, reflecting stagnation phases during which no lower-energy configurations are sampled. During such phases, the model continues to perform maximum likelihood training on the historically best samples retained in the buffer, thereby maintaining optimization pressure. Such stagnation can be partially attributed to the limited batch size employed in each sampling iteration, which constrains the model’s ability to adequately explore the parameter landscape. More fundamentally, as training progresses, the variance of the learned distribution tends to diminish, leading to over-concentration around already-discovered optima. While such contraction may enhance confidence in local predictions, it simultaneously suppresses exploration and reduces the likelihood of discovering superior solutions. This phenomenon can ultimately impede the accuracy that Flow-VQE can achieve as a standalone optimizer, thereby underscoring the necessity of incorporating diversity-preserving or exploration-enhancing sampling techniques in future developments.

\subsection{Post-training trajectories \label{app_post_training}}

\begin{figure*} 
    \centering \includegraphics[width=\textwidth]{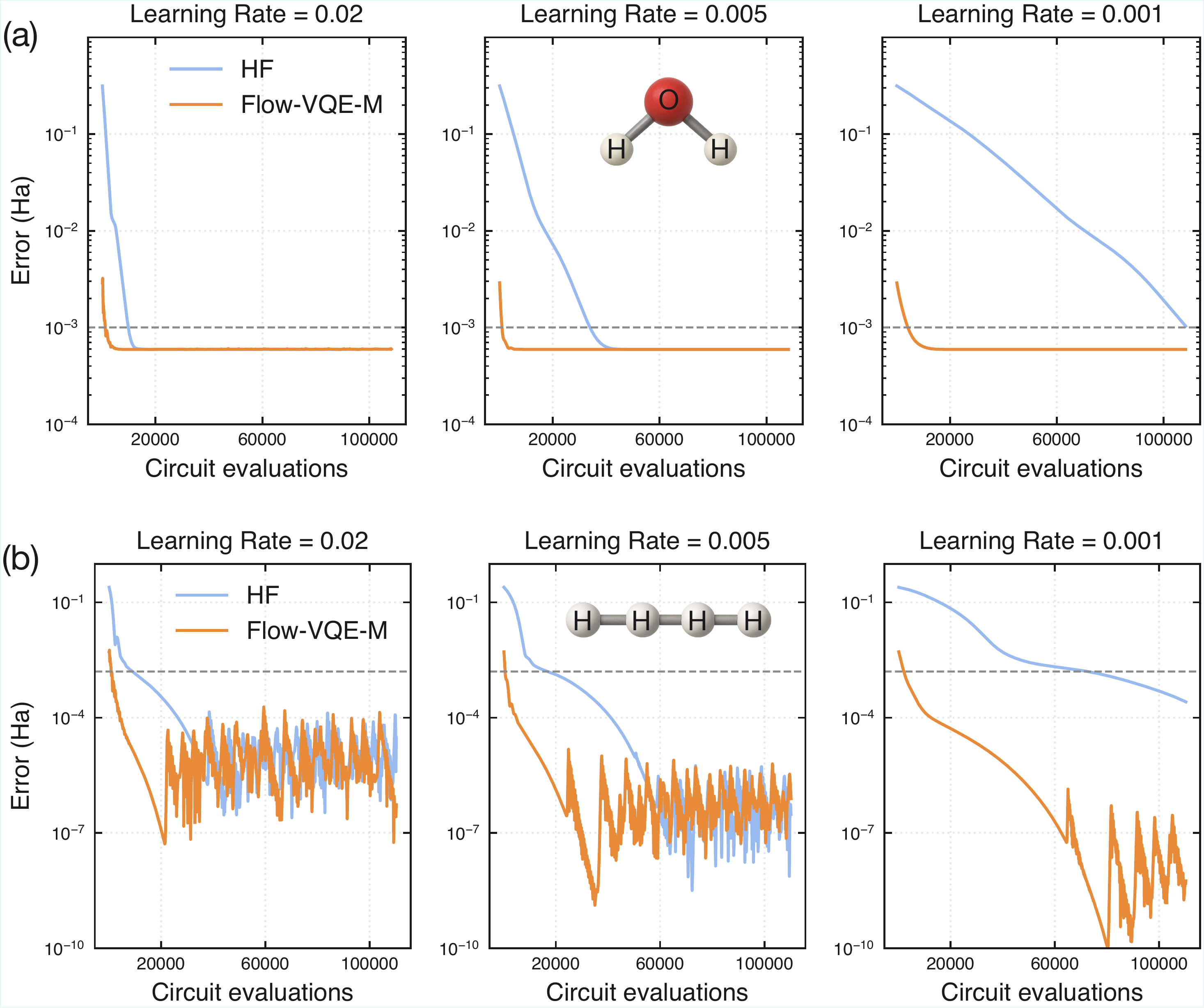}
    \caption{Post-training trajectories with initialization from HF and Flow-VQE-M, using the Adam optimizer with learning rates of 0.02 (left), 0.005 (middle), and 0.001 (right) for molecules (a) H$_2$O with $R=\SI{1.90}{\angstrom}$ and (b) H$_4$ with $R=\SI{2.58}{\angstrom}$. The gray dashed line indicates the threshold for computational accuracy.}
    \label{fig:warm_post_train_curve}
\end{figure*}

To complement the quantitative results presented as Table~\ref{tab:post_fine_tuning} in the main text, \figref{fig:warm_post_train_curve} provides the corresponding optimization curves to further elucidate the convergence dynamics when using Flow-VQE-M and HF initialization.
A key observation is the consistently steeper initial descent of Flow-VQE-M across all learning-rate settings, suggesting that the warm-started parameters lie in regions of the optimization landscape with larger gradients, thereby enabling faster initial energy minimization.
For H\textsubscript{4}, all curves exhibit noticeable oscillations in the later stages of optimization. This is primarily due to the rugged optimization landscape induced by the hardware-efficient ansatz, which interacts with the momentum component in Adam and leads to persistent fluctuations as the optimizer approaches the ground-state region. In our training runs with learning rates of 0.02 and 0.005, the number of circuit evaluations is sufficient for the optimization to settle into a quasi-stationary regime, where further improvements become trivial. Nonetheless, Flow-VQE-M consistently achieves lower energy errors than HF initialization, even in the ``bumpy'' landscape.


\section{Generated results for NH$_3$ and C$_6$H$_6$ \label{app_nh3_c6h6}}

\begin{figure*} 
    \centering 
    \includegraphics[width=\textwidth]{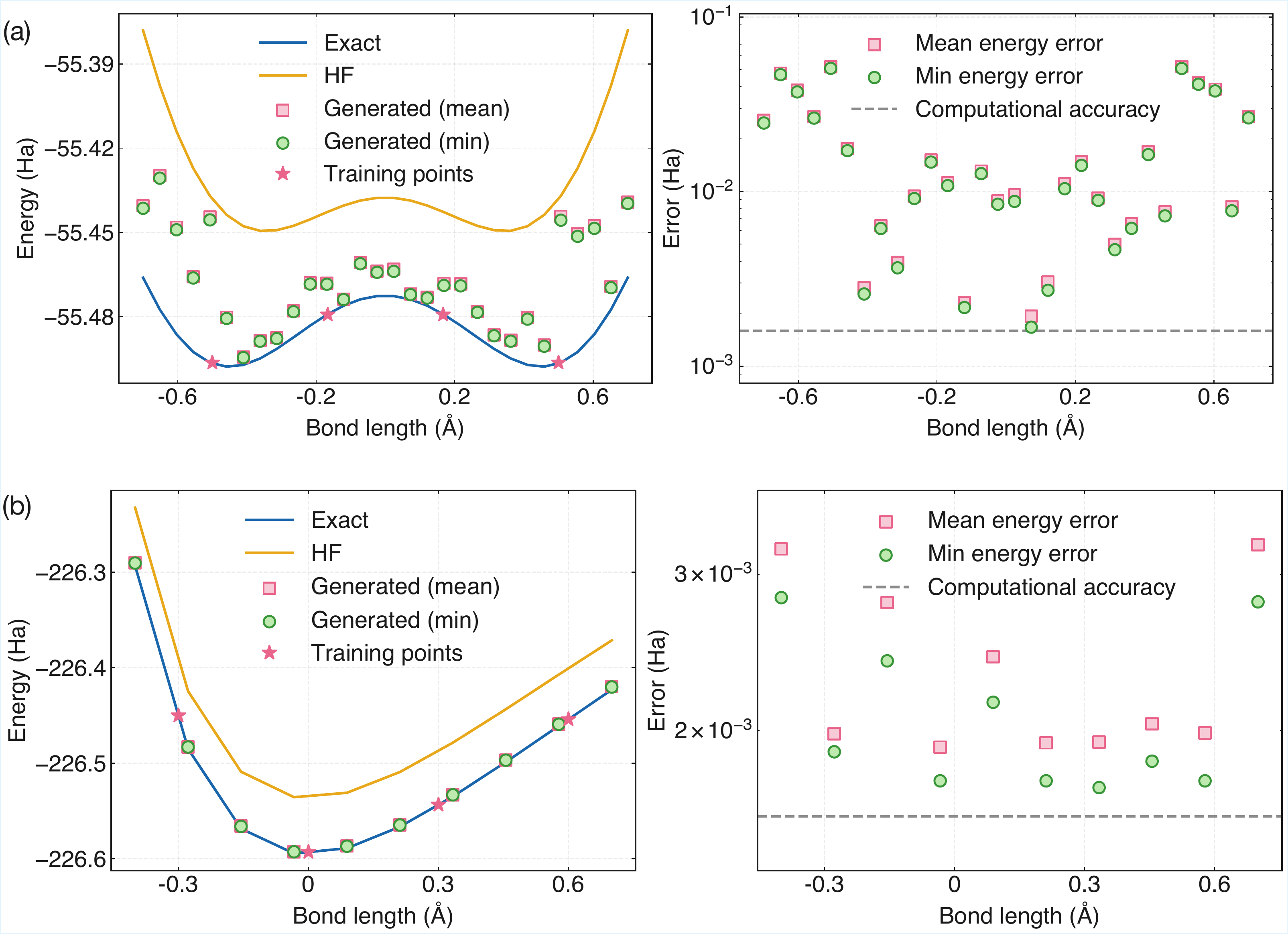}
    \caption{Potential energy surfaces (left) and corresponding errors (right) for the molecular systems (a) NH$_3$ and (b) C$_6$H$_6$ evaluated using parameters generated by Flow-VQE-M.}
    \label{fig:warm_pes_nh3_c6h6}
\end{figure*}

Figure~\ref{fig:warm_pes_nh3_c6h6} shows the PES points for NH$_3$ and C$_6$H$_6$ generated by Flow-VQE-M. Among the 16 sampled points, the one with the lowest energy is selected as the initialization corresponding to the results presented in \figref{fig:post_training_cost} of the main text. Since the training is intentionally lightweight, the generated points are not expected to reach computational accuracy at this stage.

\bibliography{reference} 
 
\newpage 

\end{document}